\documentclass[10pt,conference]{IEEEtran}

\usepackage{amsmath}
\usepackage{amsthm}
\usepackage{eucal}
\usepackage{amssymb}
\usepackage{mathrsfs}
\usepackage{times}
\usepackage{graphics}
\usepackage{graphicx}
\graphicspath{{./figures/}}
\usepackage{cite}
\usepackage{algorithmic}
\usepackage{xcolor}
\usepackage{textcomp}
\usepackage{multirow}
\usepackage{array}
\usepackage{booktabs}
\usepackage{gensymb}
\IEEEoverridecommandlockouts
\IEEEpubid{\makebox[\columnwidth]{978-1-7281-8895-9/20/\$31.00~\copyright~2022~IEEE \hfill}
\hspace{\columnsep}\makebox[\columnwidth]{ }}

\begin{document}


\title{Deep Learning for THz Channel Estimation and Beamforming Prediction via Sub-6GHz Channel}

\vspace{-10pt}

\author{
\vspace{-10pt}
\IEEEauthorblockN{Sagnik Bhattacharya, Abhishek K. Gupta}
\IEEEauthorblockA{\\Dept. of Electrical Engineering, IIT Kanpur, India\\
Emails: \{bsagnik, gkrabhi\}@iitk.ac.in} 
\vspace{-10pt}
}

\vspace{-30pt}

\maketitle

\begin{abstract}
An efficient  channel estimation is of vital importance to help THz communication systems achieve their full potential. 
Conventional uplink channel estimation methods, such as least square estimation, are practically inefficient for THz systems because of their large computation overhead. In this paper, we propose an efficient convolutional neural network (CNN) based THz channel estimator that estimates the THz channel factors  using uplink sub-6GHz channel. Further, we use the estimated THz channel factors to predict the optimal beamformer from a pre-given codebook, using a dense neural network. We not only  get rid of the overhead associated with the conventional methods, but  also achieve near-optimal spectral efficiency rates using the proposed beamformer predictor. The proposed method also outperforms deep learning based beamformer predictors accepting THz channel matrices as input, thus proving the validity and efficiency of our sub-6GHz based approach.
\end{abstract}


\section{Introduction}
\label{sec:introduction}
The ever increasing need for reliable communication with low latency and high data rates has pushed modern communication systems to utilize the mmWave spectrum \cite{THz_survey_1, THz_survey_3}. Despite their unfavourable propagation and blockage penetration characteristics, current developments in device and antenna technologies have made mmWave communication feasible, 
thus obtaining the required low latency and Gb/s data rates. 
As we move towards further-advanced wireless applications, e.g. augmented reality, and ultra-low latency video conferencing with even the Gb/s rates provided by the mmWaves falling short of the requirement, the use of THz spectrum (0.1THz-10THz) for communication is inevitable. 
THz channels suffer from an exponential pathloss decay due to severe atmospheric attenuation, along with high blockage and penetration losses \cite{THz_survey_1, THz_survey_2}. Hence, systems operating in the THz spectrum would require highly dense base station (BS) deployments. THz channels need to be highly directional in nature to compensate for the high pathloss decay, and thus require extensive beamforming with large antenna arrays at transmitter and/or receiver side(s). For such large antenna arrays, fully digital beamforming no longer remains as option, due to its high computational complexity and high power consumption \cite{THz_survey_1, THz_survey_2}. Therefore, analog beamforming and hybrid beamforming are proposed as viable solutions \cite{THz_survey_1, hybrid_precoding}. While these highly directional ``pencil beams'' improve the signal strength at the intended receiver, they often suffer from huge latency and overhead due to beam training. Not only is an accurate estimate of the THz channel required in real time, but an efficient method of choosing the optimal beamformer also needs to be devised. 

For all the reasons mentioned above, it is of fundamental importance to have a reasonably accurate estimation of the THz channel in real time. Conventional THz channel estimation methods include applying optimization such as least squared (LS) estimation, linear minimum mean squared estimation (LMMSE), etc. with the help of uplink pilot signals. But, owing to the long antenna array lengths, these estimation methods often require a large number of pilot signals. As discussed in \cite{temperature_CNN_CSI}, these methods also involve several computationally expensive matrix operations, e.g. matrix inversion, eigenvalue decomposition, singular value decomposition, etc. Compressive sensing methods have been applied for THz channel estimation\cite{compressive_sensing_1, compressive_sensing_2}. Although they reduce the time complexity compared to conventional methods, they still have significant overhead, especially considering real-time applications. Sub-6GHz channel, unlike THz channel, can be estimated with relatively low overhead, and a smaller number of uplink pilot signals. This motivates us to explore the use of the estimated sub-6GHz channel to predict the THz channel. The success of deep learning in mapping intractable functions in a time efficient manner has led to their manifold applications in wireless networks\cite{deepMIMO, deep_learning_coordinated_BS_mobile_user, viwi}. In \cite{sub6_preds_mmWave}, authors use deep learning to predict the optimal beamformer for mmWave communication, from the estimated sub-6GHz channel. In \cite{temperature_CNN_CSI}, they use various ambient factors to predict the mmWave channel state information, via a CNN. 

In this paper, we first show that there exists a mapping from the sub-6GHz channel to the THz channel factors (pathloss, angle of arrival, time of arrival, etc. for each of the channel paths). We then propose a CNN based algorithm to estimate the THz channel factors from uplink sub-6GHz channel information. The obtained channel factors can be directly used for various applications such as optimal beamformer selection, blockage prediction, etc. In this paper, we design a deep learning algorithm to directly predict the optimal beamformer by taking the CNN estimated THz channel factors as input, eliminating the overhead of obtaining the THz channel matrix. The algorithm not only achieves near-optimal spectral efficiency rates in a time efficient manner, but also outperforms a deep learning based beamformer prediction that directly uses the uplink THz channel instead.

The rest of the paper is arranged as follows. Section~\ref{sec:system-model} describes the system model for the sub-6GHz and THz scenarios. Section~\ref{sec:channel-model} describes the sub-6GHz and THz channel models. Section~\ref{sec:cnn} explains the CNN based algorithm for THz channel factors estimation using uplink sub-6GHz channel. Section~\ref{sec:nn} explains the deep learning algorithm to directly obtain optimal beamformer from the estimated THz factors. Section~\ref{sec:results} explains the obtained results and section~\ref{sec:conclusion} concludes the paper.

\section{System Model}
\label{sec:system-model}

\begin{figure}
    \centering
    \includegraphics[trim = 300 180 250 80, clip, width=\columnwidth, height = 5cm]{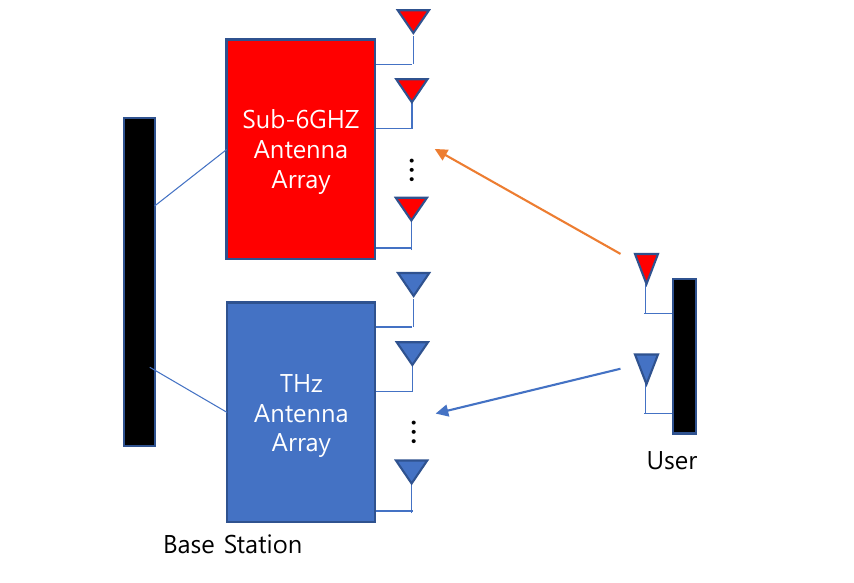}
    \caption{Illustration showing a dual-band network system  with sub-6GHz and THz antennas}
    \label{fig:system-model}
\end{figure}

In this paper, we consider a dual band network where each BS comprises two transceivers: one operating at sub-6GHz frequency with $N_{\mathrm{S}}$ antennas and the other operating at THz frequency having $N_{\mathrm{T}}$ antennas (See Fig.~\ref{fig:system-model}). For the assumption regarding the mapping from sub-6GHz channel to THz channel factors that we make later, it is required that the two transceivers are approximately colocated. The user, assumed to be a mobile device, again comprises two transceivers, corresponding to the sub-6GHz and THz frequencies, having one antenna each. We consider an OFDM system having $K$ subcarriers. The number of sub-6GHz antennas $N_{\mathrm{S}}$ being less in number, fully digital beamforming can be employed, allowing baseband channel estimation for sub-6GHz channel.

The uplink signal data is transmitted from the user to the BS via the sub-6GHz channel. Thus, at the $k$th subcarrier, for $k = 1,2,\cdots,K$, the uplink received signal is 
\begin{align}
    \mathbf{y_{\mathrm{S}}}[k] = \mathbf{h_{\mathrm{S}}}[k]~x_{\mathrm{S}}[k] + \mathbf{n_{\mathrm{S}}}[k]
\end{align}
where $\mathbf{y_{\mathrm{S}}}[k]$ is the received uplink symbol vector, $\mathbf{h_{\mathrm{S}}}[k]\in\mathbb{C}^{N_{\mathrm{S}}\times1}$ is the sub-6GHz channel vector from the user to the BS for the $k$th subcarrier, $x_{\mathrm{S}}[k]$ is the input symbol, and $\mathbf{n_{\mathrm{S}}}[k]$ is the received noise vector at the output which follows $\mathcal{N}_\mathbb{C}(0,\sigma^{2}_{\mathrm{S}}I)$. As for the downlink THz transmission, owing to the large number of antennas which makes digital beamforming computationally prohibtive and power consuming, a fully analog beamforming is adopted, with a single RF chain and $N_{\mathrm{T}}$ phase shifters. The phase shifters are quantized because of hardware constraints. Thus there is a finite set of possible beamformers, collected in the codebook $P$, each of whose elements is a candidate beamformer $p\in\mathbb{C}^{N_{\mathrm{T}}\times1}$ . Hence the downlink system model is
\begin{align}
    y_{\mathrm{T}} = \mathbf{h}^{H}_{\mathrm{T}}~\mathbf{p}~x_{\mathrm{T}} + n_{\mathrm{T}}
\end{align}
where $y_{\mathrm{T}}$ is the received output symbol at the receiver, $\mathbf{h}_{\mathrm{T}}\in\mathbb{C}^{N_{\mathrm{T}}\times1}$ is the THz channel vector from the user to the BS, $\mathbf{p}$ is the chosen beamformer from the codebook $P$, $x_{\mathrm{T}}$ is the transmit symbol, and $n_{\mathrm{T}}$ is the noise at the output following $\mathcal{N}_\mathbb{C}(0,\sigma^{2}_{\mathrm{T}})$.

\section{Channel Model}
\label{sec:channel-model}
While THz waves show propagation characteristics similar to mmWaves, they stand out in terms of the significant molecular absorption even in the free space propagation scenario \cite{THz_survey_1}. Thus, as explained in \cite{THz_survey_1}, an additional exponential term needs to be added in the pathloss equation along with the power law term. Thus for an LOS signal,
\begin{align}
    P_{\mathrm{r}} = P_{\mathrm{t}}~\lambda(r)~\delta(r)
\end{align}
where $r$ is the distance, $\lambda(r)$ is the power law pathloss, and $\delta(r)$ is the exponential term due to molecular absorption. We consider the combined pathloss $\lambda^{'}(r)=\lambda(r)\delta(r)$ as one of the THz channel factors to be estimated.

As in \cite{THz_survey_2}, considering the Doppler shifts of all the paths to be significantly small,  we adopt a geometric channel model for both the sub-6GHz and the THz channels. Therefore the channel model, for sub-6GHz channel, for the $k$th subcarrier, is given by
\begin{align}
\label{eq:channel-model}
    \mathbf{h_{\mathrm{S}}}[k] = \sum^{C-1}_{c=0}\sum^{M-1}_{m=0}\lambda^{'}_{l}e^{-j\frac{2\pi k}{K}c}~\alpha_{r}(\phi_{r},\theta_{r})\alpha^{*}_{t}(\phi_{t},\theta_{t})
\end{align}
where $C$ is the cyclic prefix length, $M$ is the number of paths, $\lambda^{'}_{l}$ is the path gain, including the effective pathloss as explained above, $\phi_{r}$ and $\theta_{r}$ are the azimuth and elevation angles of arrival respectively, $\phi_{t}$ and $\theta_{t}$ are the azimuth and elevation angles of departure respectively, $\alpha_{r}$ and $\alpha_{t}$ are the receive and transmit beamsteering vectors respectively. The THz counterpart of \eqref{eq:channel-model} is obtained by replacing the components of the equation with their THz channel factor counterparts. For the rest of the paper, whenever we mention channel factors, we mean the pathloss, phase change on arrival, time of arrival, angles of arrival and departure of each of the channel paths between a user and a BS. Thus, as computed in \eqref{eq:channel-model}, given these channel factors for every channel path, the channel is determined completely.

\section{THz Channel Factors Estimation}
\label{sec:cnn}

\begin{figure}
    \centering
    \includegraphics[trim = 80 0 240 0, clip, width=\columnwidth, height = 6cm]{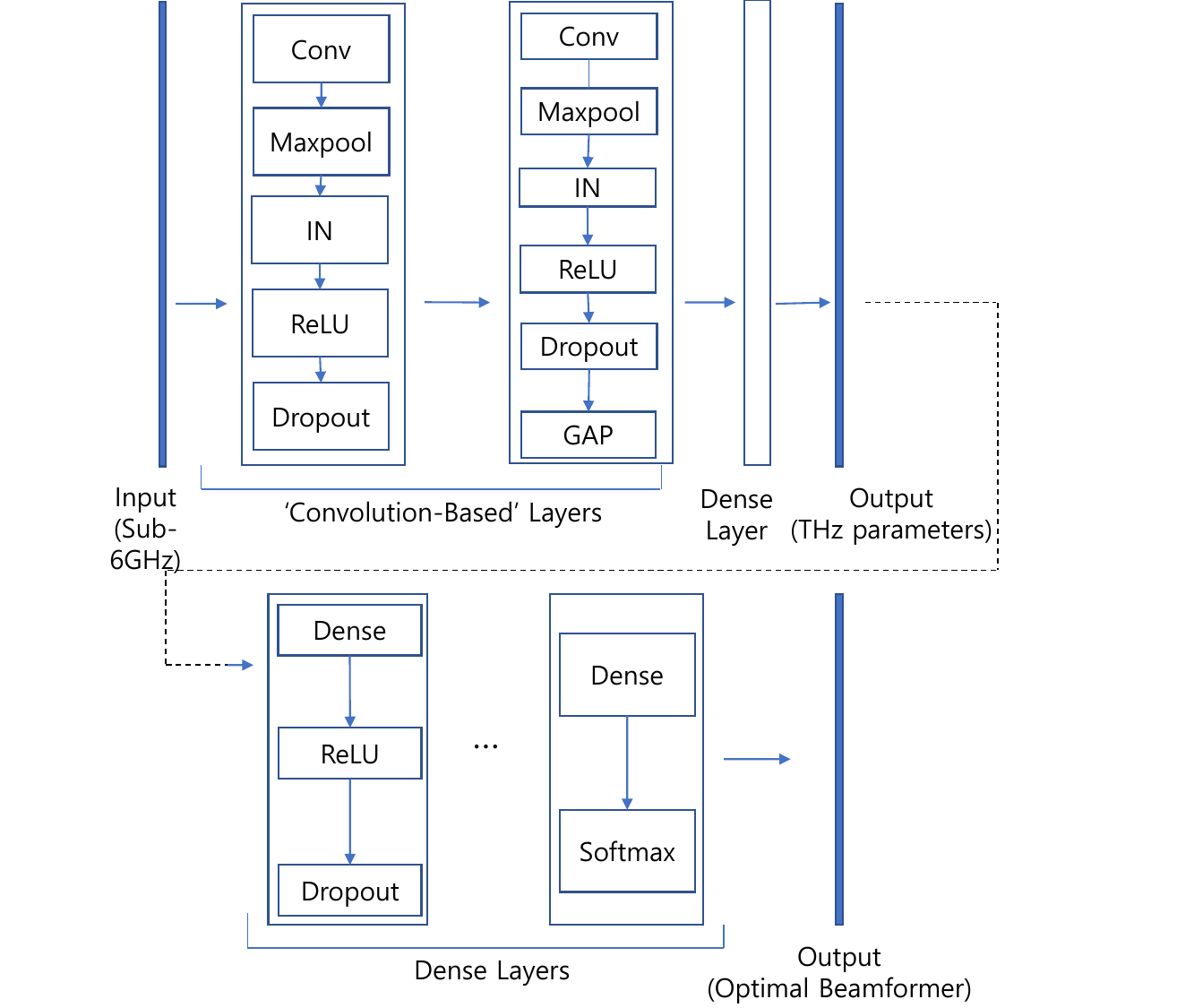}
    \caption{Architecture of (a) CNN based THz factors estimator (top) (b) Optimal Beamformer Predictor (bottom)}
    \label{fig:cnn}
\end{figure}

\textit{Assumption 1: The mapping $f$ from the channel factors to the channel matrix is bijective under suitable conditions.}

In \eqref{eq:channel-model}, it is shown that given the channel factors for every channel path, the channel matrix is determined completely. We assume that for a fixed antenna array, and surrounding environment conditions, no two set of non-identical channel factors can lead to the same channel matrix. Although this assumption depends on the environment conditions, user and BS antenna array length, array geometry, and static/dynamic characteristics, for all practical purposes, a sufficiently large number of antennas in a static environment are sufficient to ensure the bijectivity. Thus, we define the inverse function, i.e. the function from the channel matrix $\mathbf{h_{\mathrm{T}}}$ to the channel factors $\mathbf{S}_{\mathrm{T}}=[\lambda^{'}_{l},\tau,\theta,\phi_{r},\theta_{r},\phi_{t},\theta_{t} ]$, where the elements of the vector indicate the pathloss, time of arrival, phase change on arrival, azimuth and elevation angles of arrival and departure, respectively. Therefore
\begin{align}
\label{eq:THz-to-parameters}
    f^{-1}:\mathbf{h_{\mathrm{T}}}\rightarrow \mathbf{S_{\mathrm{T}}}
\end{align}
where $\mathbf{h_{\mathrm{T}}}\in\mathbb{C}^{K\times N_{\mathrm{T}}}$, and $\mathbf{S_{\mathrm{T}}}\in\mathbb{R}^{7\times 1}$.

In \cite{channel_mapping_proof}, it was shown that under certain conditions, there exists a bijective mapping from the channel between a user and a set of antennas at a particular frequency to the channel at a different frequency. Hence, given the user and the BS antennas, for the same environment and propagation conditions, there exists a function $g$ such that
\begin{align}
\label{eq:sub-6-to-THz}
    g:\mathbf{h_{\mathrm{S}}}\rightarrow\mathbf{h_{\mathrm{T}}}
\end{align}
Equations~\eqref{eq:THz-to-parameters} and \eqref{eq:sub-6-to-THz} combinedly show that there exists a mapping between the sub-6GHz channel and the THz channel factors under suitable conditions.

The above statement is exploited to derive the THz channel factors directly from the sub-6GHz channel matrix. We then use the estimated channel factors directly for a beamforming application without resorting to channel matrix computation. CNNs are extremely successful in extracting important spatial features from raw data, without requiring explicit handcrafted features. We use a CNN based network for THz channel factors estimation; this not only generates highly accurate results but also gets rid of the computational overhead associated with THz uplink based channel estimation by conventional methods, and subsequent beam training. The complete CNN based algorithm is as follows:

\textit{Input and Output : }Each element of the estimated sub-6GHz channel matrix $\mathbf{h_{\mathrm{S}}}\in\mathbb{C}^{K\times N_{\mathrm{S}}}$ is separated into magnitude and phase angle. The resulting 3D matrix $\mathbf{h^{'}_{\mathrm{S}}}\in\mathbb{R}^{K\times N_{\mathrm{S}} \times 2}$ is supplied as input (2D image format with 2 channels). The output comprises the THz channel factors for the first $L_{\mathrm{T}}$ THz channel paths, where $L_{\mathrm{T}}$ is a supplied hyperparameter. Thus the output is $\mathbf{S}_{\mathrm{T}}\in\mathbb{R}^{L_{\mathrm{T}}\times 7}$, where $\mathbf{S}_{\mathrm{T}}[l]$ contains the channel factors for the $l$th channel path, as mentioned before.

\textit{Training Phase : }In the training phase, raytracing data is obtained for both sub-6GHz and THz channels. While the sub-6GHz channel matrix is computed using the raytracing data, the raw THz channel factors are taken. These are used to generate the training samples $(\mathbf{h_{\mathrm{S}}}, \mathbf{S}_{\mathrm{T}})$, for each user, supplied to the neural network.

\textit{Algorithm Deployment Phase : }In the deployment phase, the sub-6GHz channel matrix is estimated from sub-6GHz channel uplink signals, and passed into the neural network as input. The output of the network comprises the factors for the first $L_{\mathrm{T}}$ THz channel paths. These obtained channel factors can be directly used for manifold applications without going through the THz channel matrix estimation stage.

\textit{CNN Architecture : }The top part of Fig.~\ref{fig:cnn} shows the CNN architecture adopted for this algorithm. It comprises two ``convolution-based'' layers. Each ``convolution-based'' layer comprises a convolutional layer, followed by a maxpooling, instance normalization (IN) and rectified linear unit (ReLU) activation layers. The convolutional layer uses filters of pre-specified kernel size to extract meaningful spatial features from the sub-6GHz channel matrix data. The maxpooling layer performs a maxpool operation, i.e. selecting the maximum among all the elements in the pre-specified kernel size, and helps in downsizing the input representation to reduce complexity. The IN layer, unlike conventional batch normalization (BN), applies normalization per data sample along a pre-specified axis, thus reducing incoming noise and dependence on entire dataset statistics. This improves the network performance significantly. Lastly, the ReLU activation is applied to introduce non-linearity while maintaining the range diversity of the input features.

\begin{table}[t!]
    \normalsize
    \centering
    \caption{Neural Network Hyperparameters}
    \label{tab:cnn}
    \resizebox{\columnwidth}{!}{%
    \begin{tabular}{|c|c|c|}
        \hline
        Hyperparameter & THz Estimation & Beamformer Prediction\\
        \hline
        Conv2D Kernel Size & (2,2) & -\\
        Maxpool Kernel Size & (2,2) & -\\
        No. of Training Samples & 80,000 & 80,000\\
        No. of Test Samples & 20,000 & 20,000\\
        Epochs & 100 & 100\\
        Type & Regression & Classification\\
        Optimizer & Stochastic Gradient Descent & Adam\\
        Initial Learning Rate (LR) & $1.00\times 10^{-3}$ & $1.00\times 10^{-2}$\\
        LR Decay Factor & 0.1 & -\\
        LR Decay Schedule & 80 epochs & -\\
        Momentum & 0.8 & -\\
        Dropout & 0.2 & 0.2\\
        \hline
    \end{tabular}%
    }
\end{table}

The output from the second convolution layer is passed into a global average pooling (GAP) layer, followed by a dense layer. The GAP layer computes the average of the data across all but the last dimension. Unlike a simple flattening layer, this downsizes the data dimension according to the kernel size, which helps extract feature representations while preventing overfitting. This generates faster convergence and better performance. In addition to this, a dropout layer is also introduced after each convolutional layer to randomly drop some fraction of the neurons (0.2, for our case) at algorithm runtime. This prevents the neural network from overfitting to training data. The dense layer generates the final regression output, i.e. the required THz channel factors. Table~\ref{tab:cnn} represents the network hyperparameters used for the CNN.


\section{Beamforming with Estimated THz Channel Factors}
\label{sec:nn}
We use the CNN predicted THz channel factors directly for a beamforming application. Given the sub-6GHz channel matrix information, we predict the optimal beamformer from a pre-provided codebook for downlink THz data transmission. We again use a dense neural network architecture for this classification problem (choosing the optimal beamformer from the series of available ones).

\textit{Input and Output : }The input comprises the estimated channel factors from the aforementioned CNN, i.e. the input is $S_{\mathrm{T}}$. The output is a number representing the index of the beamformer to be chosen from the given codebook. This is a categorical classification problem in its essence.

\textit{Training Phase : }During the training phase, the THz raytracing data is obtained in the form of channel factors $S_{\mathrm{T}}$, which is then used to compute $h_{\mathrm{T}}$. Then we perform an extensive beam training by checking the gain associated with each candidate beamformer $p$ in the codebook $P$. In particular, we choose the candidate beamformer $p$ such that
\begin{align}
\label{eq:beamform}
    \hat{p} = \underset{p\in P}{\arg\max} \left(\sum_{k=1}^{K} \log_{2} \left(1+\mathrm{SNR}~\left|\mathbf{h^{H}_{\mathrm{T}}}[k]~p\right|^{2}\right)\right)
\end{align}
where SNR is the pre-specified per-subcarrier transmit signal to noise ratio. Hence, we get our neural network samples as $(S_{\mathrm{T}}, p_{\mathrm{ind}})$, where $p_{\mathrm{ind}}$ is the categorical variable representing the index of beamformer $p$ in codebook $P$.

\textit{Algorithm Deployment Phase : } In the deployment phase, the estimated THz channel factors $S_{\mathrm{T}}$ from the CNN are provided as input to the network, which gives the categorical variable $p_{\mathrm{ind}}$ representing the index of the optimal beamformer from the codebook $P$.

\textit{Network Architecture : }The bottom part of Fig.~\ref{fig:cnn} shows the neural network architecture for the above prediction problem. It comprises 4 dense layers, each followed by a ReLU activation layer. Here also we add a dropout layer after each layer to reduce overfitting to training data. The output of the last dense layer is sent to a softmax layer which outputs the categorical index for the optimal beamformer.

\textit{Baseline : }For comparing the performance of our algorithm, we choose the direct THz matrix based beamformer prediction. Here the THz channel matrix is computed from raytracing data, and it is given as input into a convolutional layer, followed by global average pooling and ReLU activation. The output of the convolution layer is then passed into a dense neural network comprising 4 dense layers, each followed by ReLU and dropout layers. The above described network architecture generates the best baseline prediction performance.

\textit{Upper Bound : }The optimal upper bound (UB) performance is obtained by using the exact THz channel generated from the THz raytracing data. Then a computationally expensive exhaustive codebook survey is conducted, to select the beamformer providing the maximum gain \eqref{eq:beamform}.

\section{Performance Evaluation}
\label{sec:results}

\begin{table}
    \normalsize
    \centering
    \caption{Scenario Parameters}
    \label{tab:scenario}
    \resizebox{\columnwidth}{!}{%
    \begin{tabular}{|c|c|c|}
        \hline
        Parameter & Sub-6GHz & THz\\
        \hline
        No. of Users & 100,000 & 100,000\\
        BS Antenna Array Length & 4 & 128\\
        BS Antenna Height (m) & 8 & 8\\
        User Antenna Height (m) & 2 & 2\\
        Frequency (GHz) & 2.4 & 100\\
        Propagation Model & sbr+gas+cloud & sbr+gas+cloud\\
        Max No. of Paths & 8 & 4\\
        Bandwidth (MHz) & 20 & 500\\
        No. of OFDM Subcarriers & 64 & 64\\
        Codebook size (X,Y,Z) & $4\times 64\times 4$ & $4 \times 128\times 4$\\
        \hline
    \end{tabular}%
    }
\end{table}

\begin{figure}
    \centerline{\includegraphics[width=\columnwidth]{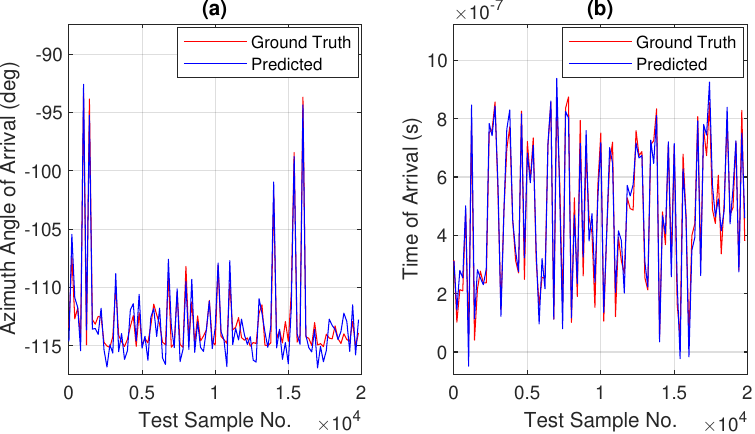}}
    \caption{Prediction performance on the first THz channel path for various channel-factors: (a) Azimuth angle of arrival (AoA) (b) Time of arrival (ToA).}
    \label{fig:THz_parameters}
\end{figure}




\begin{table}
    \normalsize
    \centering
    \caption{Prediction Error for THz Channel Factors computed across 4 Channel Paths}
    \label{tab:error}
    \resizebox{\columnwidth}{!}{%
    \begin{tabular}{|c|c|c|}
        \hline
        Average Absolute Error & Mean & Standard Deviation\\
         \hline
         AoD$~\phi$~(deg) & 5.67 & 1.23\\
         AoD~$\theta$~(deg) & 2.36 & 0.47\\
         AoA~$\phi$~(deg) & 2.22 & 0.63\\
         AoA~$\theta$~(deg) & 4.43 & 0.91\\
         Phase (deg) & 5.93 & 1.08\\
         ToA ($10^{-18} s$) & 3.80 & 0.67\\
         Pathloss ($10^{-19}$) & 5.70 & 0.35\\
         \hline
    \end{tabular}%
    }
\end{table}

\begin{figure}
    \centerline{\includegraphics[width=\columnwidth]{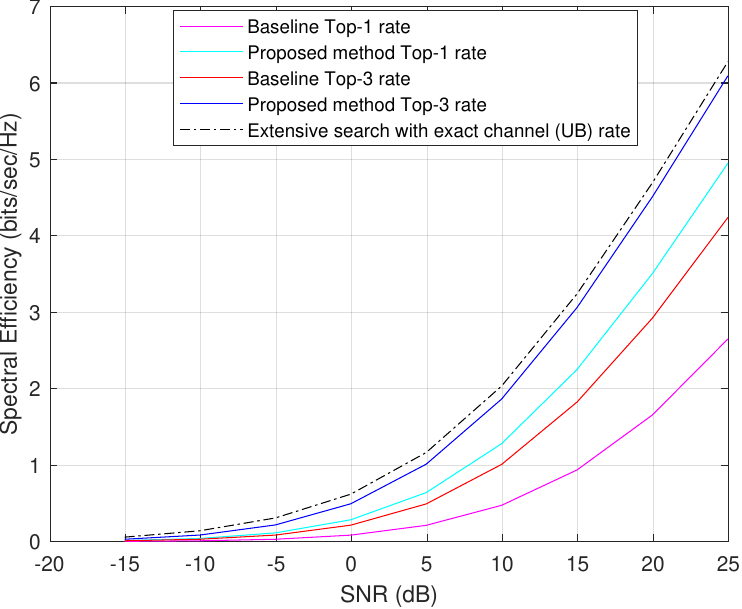}}
    \caption{Variation of the spectral efficiency using the proposed beamformer predictor with SNR. The baseline performance and the upper bound using exhaustive search are also shown for comparison.}
    \vspace{-.2in}
    \label{fig:beamforming}
\end{figure}

Table~\ref{tab:scenario} shows the parameters for the raytracing scenario generated using MATLAB raytrace command from the Communications Toolbox. Appropriate cloud and fog propagation models are added to the shooting and bouncing rays (sbr) raytracing method, to simulate the harsh atmospheric attenuation of THz waves. We choose a rectangular block of dimensions approximately $(400m\times 30m)$ in an outdoor environment, specifically the S Clarke St. of Chicago downtown area. The BS is at one of the boundaries of the rectangle and the 100,000 users are distributed randomly inside the block. The BS sub-6GHz antenna array length is taken as 4, while the THz array length is 128. We also limit the maximum number of paths from the BS to a particular user to be 8 in case of sub-6GHz case, and 4 in case of THz. 

The input to the THz factors estimation CNN is the modified sub-6GHz channel matrix $h^{'}_{\mathrm{S}}\in\mathbb{C}^{64\times 4\times 2}$ (explained in Section~\ref{sec:cnn}), generated from the uplink sub-6GHz raytracing data. The output of the CNN is $S_{\mathrm{T}}\in\mathbb{R}^{8\times 7}$. As examples, Fig.~\ref{fig:THz_parameters} shows our algorithm's prediction of the azimuth angle of arrival (AoA) and the time of arrival (ToA) of the first THz channel path, for all users. We see that the CNN is successful in predicting values close to the ground truth factors. 
Table~\ref{tab:error} shows the mean and standard deviation errors of our algorithm's THz channel factors predictions. We obtain less than 6 degree mean error on our estimates of AoA and AoD, and phase angle on arrival. Our mean errors for the ToA and Pathloss predictions are $3.80\times 10^{-18}s$ and $5.70\times 10^{-19}$ respectively, making our estimator highly accurate.

For the beam prediction network, as shown in Fig.~\ref{fig:cnn}, the estimated THz channel factors are provided as input. The output is the optimal beamformer index from the codebook containing $128\times 4\times 4 = 2048$~(quantization of phase angles along X, Y and Z axes respectively) candidate beamformers. For comparison of our algorithm, we also obtain the predicted beamformer using the baseline algorithm, as explained in Section~\ref{sec:nn}.  Fig.~\ref{fig:beamforming} shows the spectral efficiency obtained by our beamformer predictor compared to the baseline. The spectral efficiency values are derived from \eqref{eq:beamform}. The top-3 beams are obtained by taking the three maximum probability outputs of the softmax function at the last layer of the dense neural network as in Fig.~\ref{fig:cnn}. The figure contains the spectral efficiency for the top-1 predicted beam, and also the average spectral efficiency of the top-3 predicted beams. We see that our top-3 predictor achieves close to optimal (UB) rates across a wide range of SNR values $(-17\mathrm{dB}~\textrm{to}~25\mathrm{dB})$, and our top-1 predictor outperforms the top-1 and top-3 baseline predictors. The baseline THz based beamformer prediction is computationally expensive because of the larger sized $(64\times128\times2)$ channel matrix input. It is also underperforming than our sub-6GHz based predictor, because of the severe attenuation of THz signals which renders the estimated THz channel matrix highly noisy and not suitable for good learning.

\section{Conclusion}
\label{sec:conclusion}
In this paper, we implement a CNN based algorithm to estimate the THz channel factors from sub-6GHz channel input. We have then also designed a dense neural network to predict the optimal THz analog beamformer, using the CNN estimated THz channel factors. We show that our algorithm delivers better performance than direct THz uplink channel based deep learning approach, alongside being computationally cheaper. Our beamformer also achieves near-optimal spectral efficiency rates across a diverse range of transmit SNR values. The estimated THz factors can be directly used for a plethora of deep learning based THz applications, which form interesting topics for future research.

\nocite{*}
\bibliography{IEEEabrv,refs}

\end{document}